\documentclass[showpacs,amsmath,amssymb,aps,twocolumn,superscriptaddress,prx]{revtex4-2} 
\usepackage{mathtools}
\usepackage{algorithm}
\usepackage{algpseudocode}
\usepackage[pdftex]{graphicx} \graphicspath{{}}
\usepackage{grffile} 

\usepackage{comment}

\usepackage{float}
\usepackage{dcolumn}
\usepackage{bm}
\usepackage[pdftex,colorlinks=true]{hyperref}
\hypersetup{
	colorlinks=true,
	linkcolor=blue,
	urlcolor=cyan,
}

\usepackage{color}
\definecolor{ForestGreen}{RGB}{34, 139, 34}

\newcommand{\ket}[1]{\left | #1 \right\rangle}

\newcommand{\affA}{School of Physics, Peking University, 100871 Beijing, China.}
\newcommand{\affB}{Max Planck Institute for the Physics of Complex Systems, N\"othnitzer Str.~38, 01187 Dresden, Germany.}
\newcommand{\affC}{Theoretical Physics III, Center for Electronic Correlations and Magnetism,
	Institute of Physics, University of Augsburg, 86135 Augsburg, Germany.}

\begin{document}
	
	\title{Learning effective Hamiltonians for adaptive  time-evolution quantum algorithms 
	}
	\author{Hongzheng Zhao}
	\email{hzhao@pku.edu.cn}
	\affiliation{\affA}
	\affiliation{\affB}
	\author{Ao Chen}
	\affiliation{\affC}
	
	\author{Shu-Wei Liu}
	\affiliation{\affB}
	
	\author{Marin Bukov}
	\affiliation{\affB}
	
	\author{Markus Heyl}
	\affiliation{\affB}
	\affiliation{\affC}
	
	\author{Roderich Moessner}
	\affiliation{\affB}

	\date{\today}
	
	\begin{abstract}
		Digital quantum simulation {of many-body dynamics} relies on Trotterization to decompose the target time evolution into elementary quantum gates operating at a fixed equidistant time discretisation. Recent advances have outlined protocols enabling more efficient adaptive Trotter protocols, which have been shown to exhibit a controlled error in the dynamics of local observables and correlation functions. However, it has remained  open to which extent the errors on the actual generator of the dynamics, i.e., the {target many-body} Hamiltonian, remain controlled. 
		Here, we propose to use quantum Hamiltonian learning to {numerically obtain the effective Hamiltonian} and apply it on the recently introduced ADA-Trotter algorithm as a concrete demonstration. Our key observation is that deviations from the target generator remain bounded on all simulation times. This result suggests that the ADA-Trotter not only generates reliable digital quantum simulation of local dynamics, but also controllably approximates the global quantum state of the target system. Our proposal is sufficiently general and readily applicable to other adaptive time-evolution algorithms.
		
	\end{abstract}
	\maketitle
	\let\oldaddcontentsline\addcontentsline
	
	\textit{Introduction.---} Recent advances in Noisy Intermediate Scale Quantum (NISQ) devices have charted new pathways in the study of time-evolving quantum systems. Of particular interest is the
	digital quantum simulation (DQS) of many-body dynamics far away from equilibrium -- a central short-term application of quantum computation and simulation~\cite{georgescu2014quantum,daley2022practical}. Pioneering demonstrations of DQS have been achieved on various platforms, e.g., superconducting qubits~\cite{smith2019simulating,han2021experimental,guo2021observation,mi2022time,frey2022realization,keenan2022evidence}, trapped ions~\cite{lanyon2011universal,martinez2016real,zhang2017observation,monroe2021programmable}, NV centers~\cite{randall2021many,beatrez2023critical} {and Rydberg circuits~\cite{bluvstein2022quantum,bluvstein2024logical}}.
	
	DQS of real-time dynamics generated by a static target Hamiltonian $H_\ast{=}\sum_j^N H_j$ over a short time $\tau$ normally requires a decomposition of the time-evolution operator, e.g., $U(\tau) {=} \prod_{j=1}^N \exp(-i\tau H_j)$ -- a protocol known as Trotterization~\cite{suzuki1991general,berry2007efficient,poulin2014trotter,babbush2016exponentially,vanicat2018integrable,tranter2019ordering,grimsley2019adaptive,cirstoiu2020variational,barthel2020optimized,bolens2021reinforcement,lin2021real,richter2021simulating,mansuroglu2021variational,keever2022classically,tepaske2022optimal,chen2022efficient,berthusen2022quantum,mansuroglu2023problem,granet2023continuous,mansuroglu2023variational,tepaske2023compressed,trenkwalder2023compilation}. However, due to noncommutativity, $[H_i,H_j]{\neq} 0$ ($i{\neq} j$), this procedure inevitably generates so-called Trotter errors. Since the repeated application of $U(\tau)$ to simulate dynamics over a long time encodes temporal periodicity into the propagator (akin to Floquet systems), one can construct an effective Hamiltonian {$H_{\mathrm{eff}}$ perturbatively in $\tau$, such that $U(\tau){=}\exp(-i\tau H_{\mathrm{eff}})$}~\cite{goldman2014periodically,eckardt2015high,kuwahara2016floquet}. $H_{\mathrm{eff}}$ provides crucial insights to analyze Trotter errors, leading to prominent findings, e.g., rigorous bounds on Trotter errors~\cite{kuwahara2016floquet,childs2021theory,ikeda2023minimum,gong2023theory}, Trotter thresholds from localization to chaos~\cite{heyl2019quantum,sieberer2019digital,vernier2023integrable,suchsland2023krylov}, and error mitigation strategies for efficient implementation of DQS in practice~\cite{vovrosh2021simple,yang2023simulating}.
	
	Very recently there has been considerable progress in developing adaptive Trotter algorithms~\cite{wiebe2011simulating,yao2021adaptive,xu2021variational,miessen2021quantum,irmejs2023quantum,zhang2023low,zhao2023making,linteau2023adaptive,zhao2023adaptive,ikeda2023trotter24,zhao2024entanglement}. By measuring certain quantifiers of an input quantum state (such as local observables of interest or conservation laws), one can adaptively update Trotter protocols and distribute the quantum computation resources, e.g., quantum gates, more efficiently. Hence, adaptive algorithms promise to improve the performance of DQS in the NISQ era substantially. 
	Various strategies have been proposed, such as adaptively modifying the specific form of the Trotter decomposition~\cite{zhang2023low}, the step size $\tau$~\cite{zhao2023making,zhao2023adaptive}, or the variational manifold for the time-evolved quantum state~\cite{yao2021adaptive}.
	
	In this work, we raise the question of
	how to construct effective Hamiltonians for adaptive Trotter algorithms. Addressing this issue is pivotal to enhancing the stability, efficiency, and applicability of various adaptive algorithms, thereby augmenting their utility on current NISQ platforms. {
		However, identifying an effective Hamiltonian for adaptive algorithms faces key challenges. \text{(i)} Adaptive updates explicitly depend on input quantum states and one protocol optimized w.r.t. certain states may completely fail for others. This compromises the very existence of an effective Hamiltonian. \text{(ii)} Even if the latter exists, the well-established theoretical framework to perturbatively identify Trotter errors based on, e.g., Floquet-Magnus expansion which is independent of quantum states, does not apply.}
	
	Here, we aim to address these problems by harnessing the idea of Quantum Hamiltonian Learning (QHL) ~\cite{gebhart2023learning}:  {it was originally proposed to learn the Hamiltonian of a black-box quantum device; here we adapt it as a theoretical scheme to deepen our understanding of various adaptive algorithms.}
	The core idea is based on quantum state reconstruction (QSR), namely, we {numerically learn} a suitable Hamiltonian $H_{\text{eff}}^{[t]}$ such that the quantum states obtained by running the adaptive algorithm can be reconstructed from the time evolution generated by $H_{\text{eff}}^{[t]}$. Note, $H_{\text{eff}}^{[t]}$ is static but the adaptive updates make it parametrically dependent on time $t$: once $t$ is fixed, $H_{\text{eff}}^{[t]}$ is also fixed throughout the entire time evolution over the interval $[0,t]$.
	
	{Albeit intuitive, QSR is generally a resource-demanding {computational} task in practice, if there is no prior knowledge of the quantum system. This is particularly costly in a many-body setup where the optimization parameter space grows exponentially with the system size. 
		More efficient methods have been developed for static and driven systems which are not adaptive~\cite{holzapfel2015scalable,wang2017experimental,bairey2019learning,hou2019experimental,li2020hamiltonian,pastori2022characterization,nandy2023reconstructing,haah2024learning,bakshi2024structure,an2024unified,mirani2024learning,olsacher2024hamiltonian}. For instance, by measuring energy or local observables for an ensemble of initial states at various times, the Hamiltonian can be learned efficiently~\cite{bairey2019learning,li2020hamiltonian,pastori2022characterization}. Unfortunately, these methods are not applicable here since the adaptive protocol explicitly depends on both time and the input quantum states.}

\begin{figure}[t!] \includegraphics[width=\linewidth]{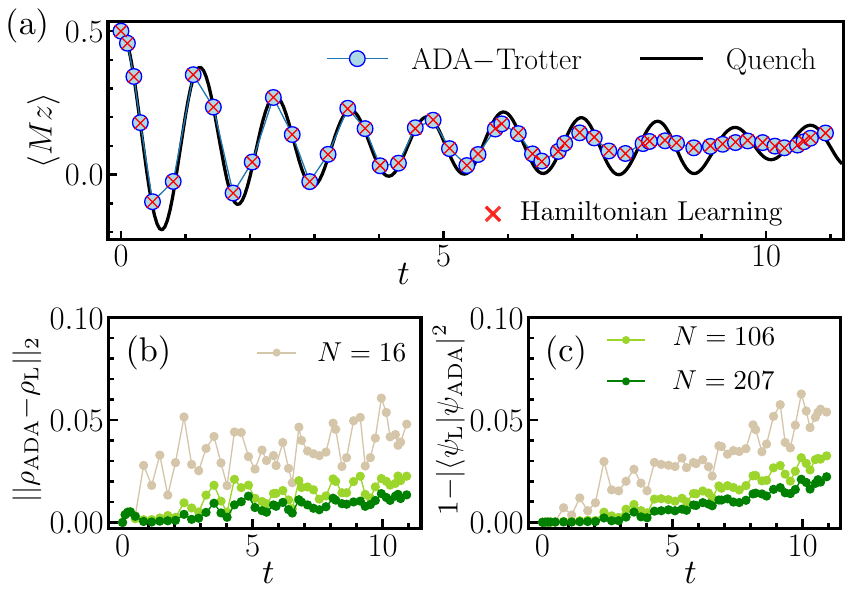}
		\caption{(a) Comparison between the magnetization of the quantum state obtained by ADA-Trotter and Hamiltonian learning. Deviations between the two in local magnetization are barely visible, implying a high-quality state reconstruction by the effective Hamiltonian. $N=207$ for the complete basis. Learning error in the three-site reduced density matrix (b) and the overall quantum state infidelity (c). Increasing the basis improves the learning outcome. A quantum
			Ising model is used with parameters $L=16, J_z=-1,h_x=-1.7,h_z=0.5$ and total Trotter steps $M=50$ for numerical simulation. The tolerances in ADA-Trotter are $(d_{\mathcal{E}},d_{\delta\mathcal{E}^2})=(0.02,0.01)$.
		}
		\label{fig.main}
	\end{figure}

	The main finding of this work is to show that for interacting systems {involving geometrically local and few-body operators}, QHL is indeed a natural tool to identify effective Hamiltonians. For a concrete demonstration, we focus on ADA-Trotter algorithm with adaptive step sizes~\cite{zhao2023making}. However, the following QHL routine is expected to be sufficiently general and readily applicable to other adaptive time-evolution algorithms. Leveraging the fact that the adaptive time evolution is dominated by a target Hamiltonian together with the intuition that most relevant Trotter errors should be sufficiently local,
	we first propose a Hamiltonian ansatz to parametrize $H_{\text{eff}}^{[t]}$. It involves terms that are local in space and obey the symmetry constraint imposed by the Trotter decomposition. We optimize it for state reconstruction via a Gradient Descent (GD) algorithm~\cite{kingma2017adam,mehta2019high}. Crucially, we find that its efficiency can be remarkably improved by recycling the optimized outcome as the starting point for the optimization onwards. This GD results in high-quality QSR for the entire time evolution, cf. Fig.~\ref{fig.main}. 
	
\begin{figure}[t]
		\includegraphics[width=0.87\linewidth]{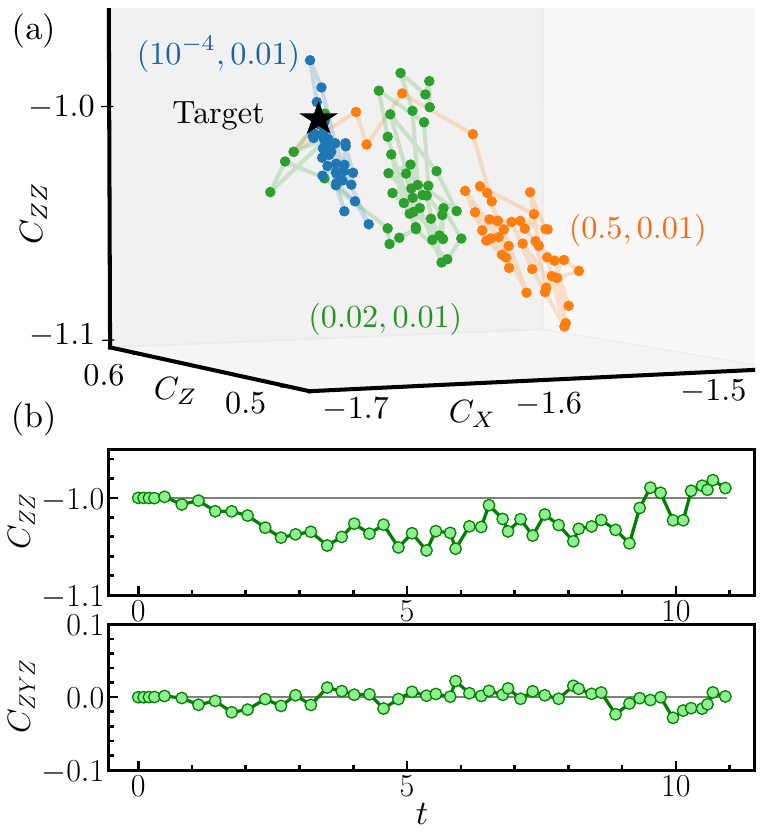}
		\caption{Learned coefficients in the effective Hamiltonian. (a) Time trace of the learned coefficients with the target highlighted as a black star. Trajectories cluster in a finite region and their deviations from the target remain bounded. ADA-Trotter thus induces an effective Hamiltonian that closely mimics the target Hamiltonian. (b) These coefficients change in time and always remain close to target values (grey). We use the same Hamiltonian parameters as in Fig.~\ref{fig.main} and tolerance values $(d_{\mathcal{E}},d_{\delta\mathcal{E}^2})$ are shown in the plot. Basis size  $N=106$ in (a) and 207 in (b).   
		}
		\label{fig.trajectory}
	\end{figure}

	QHL provides valuable insights for Trotter errors in adaptive algorithms that are not accessible otherwise. 
	As a key result, we show that although $H_{\mathrm{eff}}^{[t]}$ parametrically depends on time, it always stays close to the target Hamiltonian, cf. Fig.~\ref{fig.trajectory}. Hence, ADA-Trotter not only generates reliable DQS of local dynamics, but it also closely mimics global properties of the target many-body system. Further, by analysing the statistical behavior of errors terms, we show that leading Trotter errors derived from the Floquet-Magnus expansion still contribute notably, despite step sizes now being adaptive in time.
	
	\textit{Methods.---} We first briefly review the adaptive algorithm and elaborate on the QHL method and the associated optimization process. ADA-Trotter was first introduced in Ref.~\cite{zhao2023making} for DQS of quantum many-body dynamics generated by a target Hamiltonian $H_\ast.$ For concreteness, we consider a target non-integrable quantum Ising model, $H_\ast{=}H_z{+}H_x$, for a chain of length $L$:
	\begin{equation}
		\label{eq.example}
		H_z{=} J_z \sum\nolimits_{j} Z_{j} Z_{j+1}{+}h_z \sum\nolimits_{j} Z_j, \ \  H_x{=}h_x \sum\nolimits_{j} X_j,
	\end{equation}
	with Pauli operators $Z_j,X_j,Y_j$ and periodic boundary conditions. A second order Trotter protocol,
	$
	U(\tau_m) = e^{-i\tau_mH_x/2}e^{-i\tau_mH_z}e^{-i\tau_mH_x/2},
	$ is used here but generalization to other Trotter decompositions is straightforward. An effective Hamiltonian for a single step can be defined via the relation $U(\tau_m){:=}\exp(-i\tau_mH_{\mathrm{eff}})$. For a small step $\tau_m$, the Baker–Campbell–Hausdorff (BCH) yields
	\begin{equation}
		\label{eq.Trotter_decomposition}
		\begin{aligned}
			H_{\text{eff}} \approx H_\ast {+}\frac{\tau_m^2}{24}\left([[H_z,H_x],H_x]{+}2[[H_z,H_x],H_z]\right) ,
		\end{aligned}
	\end{equation}
	with error $\mathcal{O}(\tau_m^4)$. The leading order matches  the target Hamiltonian. ADA-Trotter automatically chooses the largest step sizes $\tau_m$ to evolve the system, as long as expectation values of energy density and variance are  preserved within preset tolerances $d_{\mathcal{E}},d_{\delta\mathcal{E}^2}$, which are the control parameters of the algorithm and set the simulation error. More precisely,
	for a given state $|\psi(t_m)\rangle$, by a feedback process we search for the largest possible time step $\tau_m$, such that, in the time-evolved state $|\psi(t_m{+}\tau_m)\rangle {=} U(\tau_m) |\psi(t_m)\rangle$,
	the energy density 
	$\mathcal{E}_{m+1} {=} L^{-1}\langle \psi(t_m{+}\tau_m) | H_\ast| \psi(t_m{+}\tau_m)  \rangle$
	and its fluctuations density
	$\delta \mathcal{E}^2_{m+1} {=} L^{-1} \langle \psi(t_m{+}\tau_m) | H_\ast^2 | \psi(t_m{+}\tau_m)\rangle - L\mathcal{E}^2_{m+1}$
	both remain  bounded:
	$
	|\mathcal{E}_{m+1}{-}\mathcal{E}| {<} d_\mathcal{E}$, $ |\delta \mathcal{E}^2_{m+1}-\delta \mathcal{E}^2| {<} d_{\delta \mathcal{E}^2},
	$
	where $\mathcal{E},\delta\mathcal{E}^2$ are energy and variance density calculated with respect to the initial state $\ket{\psi(0)}$~\cite{zhao2023making}, which evolves to $\ket{\psi_{\text{ADA}}(t)}$ at time $t$ after a few steps.
	
	The perturbative expansion in Eq.~\ref{eq.Trotter_decomposition} only works for each small time duration $\tau_m$. For two consecutive steps with different $\tau_m$ and $\tau_{m'}$, Eq.~\ref{eq.Trotter_decomposition} does not hold for the new Hamiltonian $H'_{\text{eff}}$, defined via $\exp[-i(\tau_m{+}\tau_{m'})H'_{\mathrm{eff}}]=U(\tau_m)U(\tau_{m'})$ for duration $\tau_m{+}\tau_{m'}$. Our central goal is to identify one appropriate effective Hamiltonian $H_{\text{eff}}^{[t]}$, such that $\ket{\psi_{\text{ADA}}(t)}$ can be reconstructed by the effective time evolution, $\ket{\psi_{\text{L}}(t)}{=}\exp(-itH_{\text{eff}}^{[t]})\ket{{\psi}(0)}$ from the same initial state. This method avoids the resource demanding quantum process tomography, hence permitting more efficient QHL for large system sizes. 
	
	We now assume that the effective Hamiltonian has the following form,
	$
	H_{\text{eff}}^{[t]}{=}\sum_{\alpha}C^{[t]}_{\alpha}\mathcal{O}_{\alpha},
	$
	where $\alpha$ denotes a Pauli string, and $\mathcal{O}_{\alpha}$ represents the corresponding tensor product of Pauli matrices. For instance, the coefficient for the term $\sum_{i}Z_iZ_{i+1}$ is $C_{ZZ}$ and from now on we will drop the parametric dependence on time for simplicity. Therefore, QHL is recast in identifying the most relevant terms in such an ansatz. 
	However, the number of all possible terms (even after fully exploiting the translation and inversion symmetry of the system) on neighbouring $R$ sites scales exponentially in the spatial support $R$. Hence, determining these coefficients even for systems of small size can still be resource-demanding.
	Although Eq.~(\ref{eq.Trotter_decomposition}) does not apply straightforwardly in defining $H_{\text{eff}}$, it actually indicates that Trotter errors are generated locally in space, via the form of nested commutators of $H_x$ and $H_z$. Hence, a substantial simplification for QHL can be made by truncating the ansatz to all possible terms in a finite and small support $R$. In the following, we choose $R=5$, corresponding to a $N=207$ operator basis, which is sufficiently large to capture the most relevant local Trotter errors while sufficiently small to efficiently implement the GD algorithm to be introduced below.
	
	For perfect construction of $H_{\text{eff}}$,  the state infidelity $(1 {-} |\langle \psi_{\text{ADA}}(t)\ket{\psi_{\text{L}}(t)}|^2)$ vanishes. Therefore, it provides a natural loss function to quantify the \textit{learning error} and optimize over the possible parameter space by a GD algorithm. Importantly, since the step sizes are generally smaller than the local energy scale of the target system, we expect that $C_{\alpha}$ does not change abruptly between two neighboring steps. Therefore, by using the learned coefficients obtained at one previous time step as the starting optimization point for the next time step, one can significantly improve the GD efficiency. For the first step, we set the starting optimization point as $(C_{X},C_{Z},C_{ZZ}){=}(h_x,h_z,J_{zz})$ -- the same as in the target Hamiltonian -- and all other coefficients are zero. More details about our GD implementation are given in SM.
	
	Generally the solution for $H_{\text{eff}}$ identified via QSR is not unique, simply because terms in the Hamiltonian {which have negligible projection onto the initial states}
	barely cause any dynamical effects. During the optimization process, they do not make much of a difference to the loss function, hence GD in effect concentrates on terms that actually dominate Trotter errors.
	
	\textit{Results.---}
	We now numerically implement the ADA-Trotter algorithm and confirm that GD can reach a sufficiently small learning error. Then we will discuss the behavior of $C_{\alpha}$ and its implication for ADA-Trotter. 
	
	In Fig.~\ref{fig.main}(a) we plot the dynamics of the local magnetization $M_z{=}\sum_jZ_j$ (black line) computed for the exact target system, starting from the initial state $\exp(-i\theta_y\sum_jY_j)|\downarrow\dots\downarrow\rangle$ with $\theta_y{=}\pi/3$~\footnote{More numerical results with different initial states are shown in the SM.}. $M_z$ oscillates coherently at early times and  the {oscillation} amplitude gradually decays as is typical for many-body systems due to quantum thermalization. Due to finite size effects, even at long times these oscillations persist but their amplitude is expected to vanish for $L{\to}\infty$. As shown by the blue dots, ADA-Trotter closely mimics the exact dynamics, and deviations only become noticeable at long times ($t{\geq} 8$). In fact, these deviations can be systematically suppressed by using smaller tolerances of the algorithm. 
	
	We now quantify the accuracy of QSR by analyzing the dynamics of $M_z$, reduced density matrix, and quantum state infidelity. In Fig.~\ref{fig.main}(a), we plot the reconstructed local dynamics as red crosses, and its deviation from ADA-Trotter (blue dots) is barely visible with a naked eye. 
	The two-norm of the difference between two reduced density matrix on three consecutive sites ($\rho_{\mathrm{ADA}}$ for ADA-Trotter and $\rho_{\mathrm{L}}$ for the reconstructed state) is depicted in Fig.~\ref{fig.main}(b). With the largest learning basis (dark green, $N{=}207$), their deviations increase in time but only at a very slow pace. Similar behavior also occurs in the state infidelity as shown in Fig.~\ref{fig.main}(c): even at long times $t\sim 10$ where quantum thermalization effectively already occurs, the state infidelity still remains below 0.02. All of these evidences ensure that the
	learning error is sufficiently low and hence the
	construction of $H_{\text{eff}}$ is reliable.

	As GD is time-consuming for a large operator basis set, we illustrate the possibility of truncating the basis size $N$ for the improvement of the optimization efficiency, while still maintaining a high learning quality. One possible measure to quantify the importance of a Pauli operator in contributing to Trotter errors is the deviation ($\Delta C_{\alpha}$) between the learned coefficients and their target values, averaged over both time and different tolerances when running ADA-Trotter. In SM, we sort out the operator basis and obtain an importance hierarchy. 
	We truncate the operator basis set accordingly by only involving the largest $N$ terms and again perform GD. As shown in Fig.~\ref{fig.main}(b) and (c), $N{=}16$ already leads to a decent learning quality, e.g., around $t{=}10$ the loss function can be optimized down to 0.05. Increasing $N$ generally improves the optimization outcome and requires longer optimization time, cf.\ SM for additional discussion.
	
	\begin{figure}[t]
		\includegraphics[width=\linewidth]{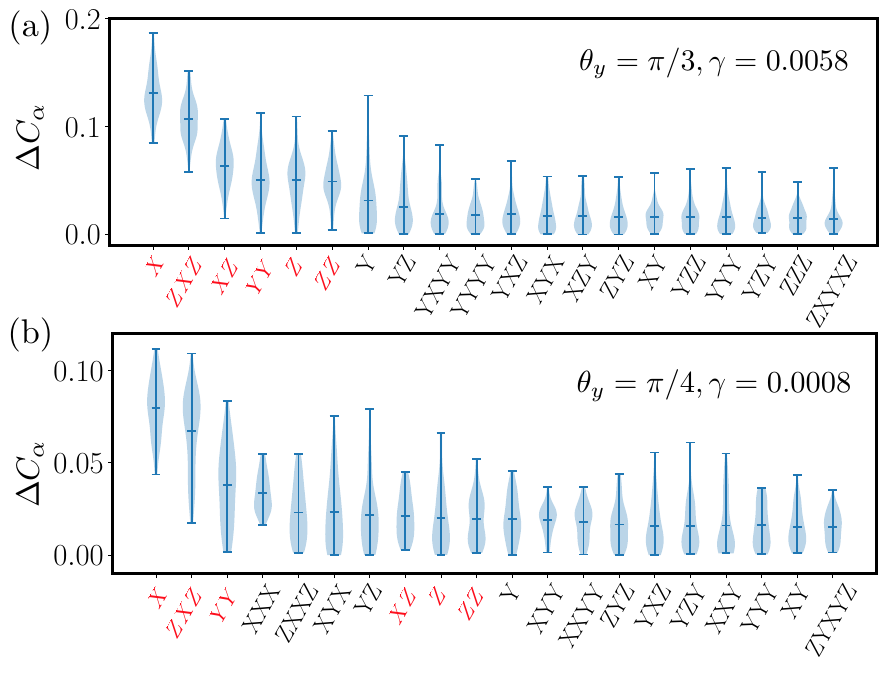}
		\caption{Violin plots for the learned coefficients after the first 10 steps for two different initial states. Leading order terms in the BCH expansion (highlighted in red) dominate the Trotter errors for adaptive algorithms. $\gamma$ stands for the increasing rate of the state infidelity, which remains low for these two cases suggesting high-quality QSR, cf. SM. We use the same Hamiltonian parameters as in Fig.~\ref{fig.main}, $(d_{\mathcal{E}},d_{\delta\mathcal{E}^2})=(0.5,0.01),N=106$ for (a) and $(d_{\mathcal{E}},d_{\delta\mathcal{E}^2})=(0.02,0.01),N=207$ for (b).
		}
		\label{fig.Violin_main}
	\end{figure}
	
	\begin{figure}[t]
		\includegraphics[width=\linewidth]{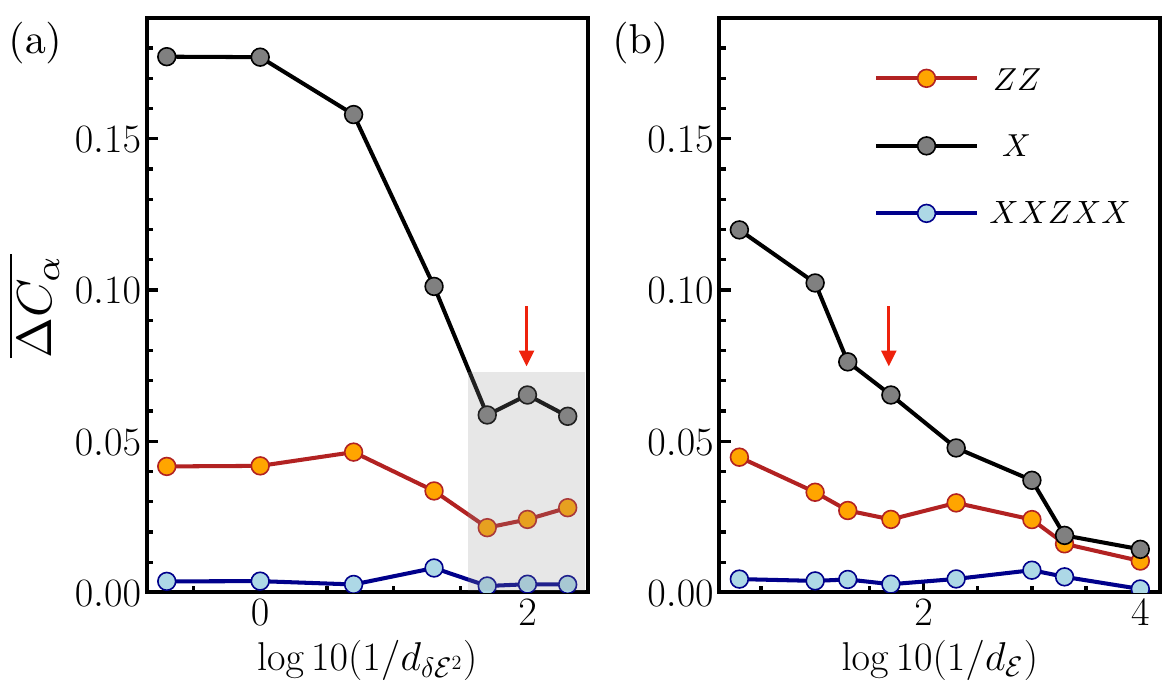}
		\caption{Dependence of a time-averaged error in ADA-Trotter algorithm on the tolerances in energy variance $d_{\delta \mathcal{E}^2}$ (a) and energy $d_{\mathcal{E}}$ (b). This error saturates in the shaded region with small $d_{\delta \mathcal{E}^2}$, but it can be further suppressed by a smaller $d_{\mathcal{E}}$. 
			The red arrow points to the data obtained by the same set of tolerance values. We fix $d_{\mathcal{E}}{=}0.02$ and $d_{\delta \mathcal{E}^2}{=}0.01$ in (a) and (b), respectively. Hamiltonian parameters are the same as in Fig.~\ref{fig.main} and $M{=}100$, $N{=}106$. 
		}
		\label{fig.epsilon}
	\end{figure}
	
	With this QHL scheme we can explicitly construct $H_{\text{eff}}$, providing valuable insights for Trotter errors in ADA-Trotter.
	In Fig.~\ref{fig.trajectory}(a), we plot the coefficient trajectories of the three most important terms ($C_{X},C_{Z},C_{ZZ}$) which are non-vanishing in $H_\ast$, for three different tolerance values. Initially, at $t=0$, no optimization is needed and we set $(C_{X},C_{Z},C_{ZZ})=(h_x,h_z,J_{zz})$ (black star) as the origin of these trajectories. At short times the state $\ket{\psi_{\text{ADA}}(t)}$ is close to the initial state, so GD quickly converges and identifies the suitable parameters for QSR. 
	The trajectory starts deviating  from the target after a short transient period. However, crucially, at later times and for all tolerance values, trajectories cluster in a finite region and their deviations from $H_\ast$ remain bounded throughout the entire time evolution. In Fig.~\ref{fig.trajectory} (b), we also plot the time trace of $C_{ZZ}$  which always stays close to its target value (grey line). For terms that do not appear in $H_\ast$, e.g., $\sum_{i}Z_iY_{i+1}Z_{i+2}$, its coefficient indeed oscillates around zero.
	Therefore, we confirm that $H_{\text{eff}}$ for ADA-Trotter closely mimics the target Hamiltonian. 
	
	{Interestingly, although the ADA-Trotter violates the temporal periodicity, terms appearing in the leading order BCH expansion, including $X,ZXZ,XZ,YY,Z,ZZ$ (cf. SM), still dominate Trotter errors. To see this, in Fig.~\ref{fig.Violin_main}, we depict a collection of violin plots for different learned coefficients $C_{\alpha}$ where the Pauli strings corresponding to BCH expansion are highlighted in red. At the middle of each violin plot the time-averaged value ($\overline{\Delta C_{\alpha}}$) is shown, and the cap corresponds to their extreme values. The overall shape of the violin plot is a kernal density plot to illustrate the probability distribution of $C_{\alpha}$. For clarity we only plot 20 terms with
		large $\overline{\Delta C_{\alpha}}$. Although certain errors can exhibit large extrem values, e.g., $\Delta C_{Y}$ in Fig.~\ref{fig.Violin_main}(a), statistically those six terms notably contributes to Trotter errors.}

	Trotter errors strongly depend on the tolerances preset in the adaptive algorithm. For instance, as shown in Fig.~\ref{fig.main}(d), trajectory clusters drift further away for larger tolerances. This occurs simply because larger tolerances generally encourage ADA-Trotter to choose larger step sizes and hence cause more simulation errors. With QHL one can now quantify such a dependence by using 
	$\overline{\Delta C_{\alpha}}$. In Fig.~\ref{fig.epsilon}, we consider three different terms and show this quantity for various energy variance tolerances in panel (a) and energy tolerances in panel (b). We note that deviations in both $C_{ZZ}$ (orange) and $C_X$ (grey) are sensitive to the variation of the tolerance values, and a tighter tolerance generally suppresses the error. In contrast, terms of a larger support can have relatively smaller errors, for instance, $C_{XXZXX}$ is barely noticeable and always remains close to zero regardless of the tolerances. In the shaded region in Fig.~\ref{fig.epsilon}(a), $\overline{\Delta C_{\alpha}}$ shows little dependence on the tolerance in energy variance, indicating that there errors are mostly induced by deviations from the energy conservation. Indeed in Fig.~\ref{fig.epsilon}(b) when we fix tolerance in energy variance, errors can be more effectively controlled by tightening the energy tolerance $d_{\mathcal{E}}$. 
	
	\textit{Discussion.---}
	{Here, we use QHL to identify a suitable $H_{\text{eff}}$ for adaptive time-evolution algorithms and apply it to ADA-Trotter as a concrete demonstration. Although ADA-Trotter only approximately conserves the lowest two moments of $H_\ast$, the precise reconstruction of $H_{\mathrm{eff}}$ suggests that the generator of the entire time evolution remains close to $H_\ast$. Their deviation is also controllable upon reducing tolerances. Therefore, ADA-Trotter generates reliable and controlled many-body states close to the target. For other adaptive algorithms, this property is not guaranteed and errors in $H_{\text{eff}}$ may diverge in time. }
	
	{The efficiency in QHL relies on the \textit{a priori} intuition of target systems and Trotter errors. The local structure with symmetries in $H_\ast$, together with the simple Trotter decomposition, greatly simplifies the parametrization of the ansatz. 
		For more general cases, where systems involve non-local terms or circuit decomposition adapts variationally, a natural choice to construct operator basis may not exist. If so, one may consider Hamiltonian structure learning~\cite{bakshi2024structure} or machine learning-assisted algorithms~\cite{granade2012robust,gentile2021learning,flynn2022quantum,nandy2023reconstructing} to make QHL efficient.}
	
	It is worth emphasising that Trotter error analysis should explicitly incorporate information of input quantum states and microscopic details of the adaptive protocol. This is particularly challenging for quantum many-body systems where normally only 
	upper bounds on errors derived for worst-case scenarios or asymptotic-in-time local errors can be obtained~\cite{childs2021theory,zhao2023making}. Our method thus can systematically benchmark different adaptive algorithms from the effective Hamiltonian perspective on an equal footing. Indeed, not being restricted to adaptive algorithms,
	it also applies to other time-dependent protocols like aperiodically driven systems where effective Hamiltonians may also be difficult to obtain~\cite{else2020long,mori2021rigorous}.

	{Finally, let us point out a connection between our approach and certain concepts in the field of shortcuts to adiabaticity~\cite{chen2010fast}. Notice that the states visited by ADA-Trotter can be thought of as a trajectory in Hilbert space; in this sense, QHL can be viewed as the problem of identifying a parent Hamiltonian which generates a dynamics that traces the same trajectory. For many-body states allowing few-parameter parameterizations, parent Hamiltonians can be reconstructed efficiently using tensor networks~\cite{ljubotina2022optimal} under certain conditions~\cite{fernandez2015frustration}, and the variational principle for adiabatic gauge potentials in counterdiabatic driving~\cite{kolodrubetz2017geometry,ljubotina2022optimal}. It is an interesting open problem how to combine these techniques with QHL for trajectories of less structured states.}
	
	\textit{Acknowledgments.---}
	{We thank Peter Zoller for initiating this work and stimulating discussions. We also thank Tobias Olsacher and Christian Kokail for many useful discussions in the beginning of this project.} This work is supported by “The Fundamental Research Funds for the Central Universities, Peking University”, and by "High-performance Computing Platform of Peking University" and by the Deutsche Forschungsgemeinschaft under the cluster of excellence ct.qmat (EXC 2147, project-id 390858490).
	This project has received funding from the European Research Council (ERC) under the European Union’s Horizon 2020 research and innovation programme (grant agreement No. 853443).
	This work was supported by the German Research Foundation DFG via project 499180199 (FOR 5522).
	Funded by the European Union (ERC, QuSimCtrl, 101113633). Views and opinions expressed are however those of the authors only and do not necessarily reflect those of the European Union or the European Research Council Executive Agency. Neither the European Union nor the granting authority can be held responsible for them.

	\bibliography{Trotter}
	\let\addcontentsline\oldaddcontentsline
	\cleardoublepage
	\onecolumngrid
	\begin{center}
		\textbf{\large{\textit{Supplementary Material} \\ \smallskip
			Learning effective Hamiltonians for adaptive  time-evolution quantum algorithms }}\\
		\hfill \break
		\smallskip
	\end{center}
	
	\renewcommand{\thefigure}{S\arabic{figure}}
	\setcounter{figure}{0}
	\renewcommand{\theequation}{S.\arabic{equation}}
	\setcounter{equation}{0}
	\renewcommand{\thesection}{SM\;\arabic{section}}
	\setcounter{section}{0}
	\tableofcontents
	
	\section{Technical details for QHL}
	\subsection{Construction of basis}
	In this section we explain how to choose an operator basis for the effective Hamiltonian ansatz. In the main text, we consider the target Hamiltonian Eq.~\ref{eq.example}, which has both the spatial translation symmetry and mirror symmetry. Since the operators $\mathcal{O}_{\alpha}$ in the ansatz Hamiltonian should originate from the nested commutators of $H_x$ and $H_z$, they also obey these symmetries. On a given site $j$, there are three possible Pauli operators $X_j,Y_j,Z_j$ and to obey the symmetry constraint one can only allow operators that uniformly sum over all sites, e.g., $\mathcal{O}_X=\sum_{j}X_j$ with the corresponding coefficient $C_{X}$. For two-sites and multi-sites operators, two possible cases can occur: First, each individual operator are  already mirror symmetric w.r.t the center of the support, e.g. $X_jX_{j+1}$ and $X_jZ_{j+1}X_{j+2}$, we use $\mathcal{O}_{XX}=\sum_j X_jX_{j+1}$ and $\mathcal{O}_{XZX}=\sum_j X_jZ_{j+1}X_{j+2}$; Second, each term is not individually mirror symmetric, e.g. $X_jY_{j+1}$, we use $\mathcal{O}_{XY}=\sum_j X_jY_{j+1}+Y_jX_{j+1}$ such that after summing over all sites the operator satisfies the symmetry constraint. 
	\subsection{Details for the Gradient Descent algorithm}
	We define the loss function as 
	\begin{equation}
		l = 1 - |\langle \psi_{\text{ADA}}(t)\ket{\psi_{\text{L}}(t)}|^2
	\end{equation}
	the infidelity between the two states, such that a small loss function indicates a high-quality of QSR. The parameters to be optimized are the coefficients $C_\alpha$ in 
	$
	H_{\text{eff}}^{[t]}{=}\sum_{\alpha}C^{[t]}_{\alpha}\mathcal{O}_{\alpha},
	$
	and they enter the loss function via $\ket{\psi_{\text{L}}(t)}{=}\exp(-itH_{\text{eff}}^{[t]})\ket{{\psi}(0)}$. 
	Building with the Optax library for Google JAX, we implement an adaptive moment estimation (Adam)\cite{kingma2017adam} in the optimization algorithm. The key difference between Adam and standard stochastic Gradient Descen (GD) is that Adam computes individual learning rates (i.e. the step size $\alpha$ in each epoch of the GD) adaptively for each parameter based on the estimated first and second moments of the gradients, while in standard stochastic GD a single learning rating is maintained for all parameter updates. {We find this feature of Adam particularly suitable for our optimization task because the loss function landscape in the coefficients space can be very complicated, and the appropriate learning rate in each epoch may vary notably throughout the optimization process. }  
	In our numerical simulation, the initial learning rate is set to be $\alpha=10^{-5}$. The concept of momentum in parameter optimization is incorporated by two hyperparameters $\beta_1$ and $\beta_2$ which are exponential decay rates for the first and second moment estimates while taking the exponential moving average. Specifically, $\beta_1$ controls how much previous gradients take part in the new update, and $\beta_2$ is used in scaling the learning rate for each parameter; parameters with a high second moment (i.e. parameters whose gradients are large and varying) receive smaller updates and vice versa. Because such average is involved, another parameter $\epsilon$ is put in place to prevent any division-by-zero error. {We use $\beta_1=0.9,\beta_2=0.99,\epsilon=10^{-8}$, default values as in JAX, but these hyperparameters typically require very little tuning.} In order to ensure efficient training, we also implement a cut-off threshold ($l_{\mathrm{min}}=10^{-4}$) such that the optimization process is terminated once the the loss function drops below this threshold. The maximal number of training epochs we set is 5000. Indeed, at early times, for instance, within five Trotter steps, $l_{\mathrm{min}}$ can be achieved quickly so we can terminate the optimization with a few number of epochs. For longer times, as shown in Fig.~\ref{fig.main}, we cannot achieve this low loss function cut-off and 5000 epochs are fully exploited.
	
	At a given time, the optimization process starts from one initial guess of the coefficients. For the first Trotter step, we use $(C_{X},C_{Z},C_{ZZ}){=}(h_x,h_z,J_{zz})$ -- the same as in the target Hamiltonian -- and all other coefficients start from zero. Later, we pass the optimized parameters as the initial guess for the next training, which notably improves the optimization efficiency.

	{A possible numerical issue in GD is the accidental degeneracy encountered when evaluating the  gradients of loss}. In the computation of gradients, one intermediate quantity is $\partial \ket{\psi_{\text{L}}} /\, \partial C_\alpha$, which can be viewed as the changing rate of the outcome state $\ket{\psi_{\text{L}}}$ under a perturbation of the Hamiltonian $H_{\text{eff}}$. When the Hamiltonian has two identical eigenvalues up to machine precision, this becomes degenerate perturbation and a naive gradient will fail. Whenever this problem happens, a small change is added to the coefficients $C_\alpha$ to lift the system away from numerical degeneracy.
	
	\section{Reduction of the operator basis}
	GD in a large parameter space can be time-consuming and it is possible to truncate the operator basis to speed up the optimization procedure. Of course, one can use the complete operator basis corresponding to a small support $R$. Here we illustrate another method to truncate the basis while still keeping some operators with a large support. The central idea is to perform QHL for many realizations of ADA-Trotter, and then we try to sort all operator basis according to their averaged coefficients deviations.
	
	\begin{figure*}[h]
		\includegraphics[width=0.5\linewidth]{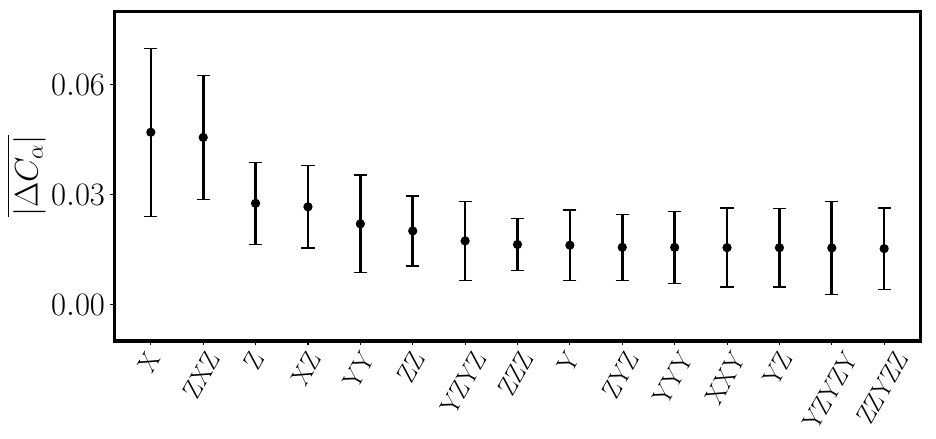}
		\caption{Tolerance-time-averaged deviations in the effective Hamiltonian for each individual term. For clarity we only plot the largest 15 terms. Few terms are notably larger than others, e.g. $X$ or $XZX$ and most of the averaged values are very small and comparable to each other. Error bars correspond to their standard deviations among the data obtained for different tolerance values. The following Hamiltonian parameters are used for ADA-Trotter $J_z=-1,h_x=-1.7,h_z=0.5,M=100$, and multiple sets of tolerances are used $d_{\mathcal{E}} = \{0.02,0.01,0.005,0.002,0.001,0.0005,0.0001\}$ and $d_{\delta \mathcal{E}^2}=\{0.02,0.03\}$; $d_{\mathcal{E}} = \{0.01,0.02\}$ and $d_{\delta \mathcal{E}^2}=\{0.05, 0.04, 0.03, 0.02, 0.01, 0.005\}$. 
		}
		\label{fig.StatisticsTolerances}
	\end{figure*}
	
	To do so, we still use the same initial state as in the main text $\exp(-i\theta_y\sum_jY_j)|\downarrow{\dots}\downarrow\rangle$ with $\theta_y=\pi/3$ and implement ADA-Trotter as well as QHL for $L=10$. Then we compute the absolute value of $C^{[t]}_{\alpha}$ subtracted by its target value and average it over time, as an indicator for the time-averaged errors in the effective Hamiltonian. We further average this time-averaged error over different realizations of ADA-Trotter given various tolerances, and sort tolerance-time-averaged errors as shown in Fig.~\ref{fig.StatisticsTolerances}. The error bar denotes
	the standard deviation among the data for different tolerances. Note, for clarity we only show the largest 15 terms out of the full operator basis with $207$ terms for $R=5$. Apart from very few terms that are notably larger than others, e.g. $X$ or $XZX$, it is generally very difficult to distinguish error terms that are dominating. Most of the averaged values are very small and comparable to each other. We truncate the operator basis by only including the largest $N$ terms and in the following we discuss the performance of GD with the truncated basis.
	
	\subsection{Performance of truncated basis---basis size $N$}
	\begin{figure}[h]
		\includegraphics[width=0.7\linewidth]{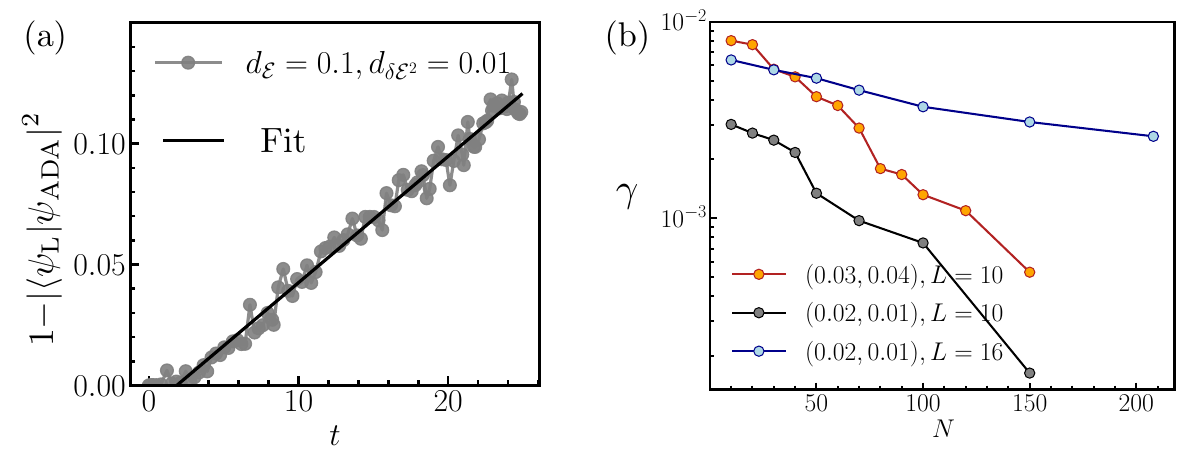}
		\caption{(a) Grey: Loss function after optimization of GD at each time. Its growth can be well fitted with a straight line (black) of slope $\gamma$. Note, the fit is only performed after the first five Trotter steps to avoid the impact of the early-time state reconstruction that is almost perfect. (b) Performance of the GD with a truncated operator basis. A larger $N$ enlarges the optimization parameter space, hence generally improves QHL and reducing the slope.
			We use $L=16, J_z=-1,h_x=-1.7,h_z=0.5,$ $(d_{\mathcal{E}},d_{\delta\mathcal{E}^2})$ are noted in the legend.  
		}
		\label{fig.slope}
	\end{figure}
	
	As already discussed in the main text, for $N=16$ GD can already achieve a sufficiently good quantum state reconstruction. Here we supply further analysis based on the growing rate $\gamma$ of the loss function after optimization at each time. As shown in  Fig.~\ref{fig.slope} (a), at very early times, since the time-evolved state $\ket{\psi_{\text{ADA}}(t)}$ remains close to the initial state, GD can always reconstruct the state up to the highest accuracy (loss value $\sim 10^{-4}$) that is preset in GD algorithm. 
	Then the state infidelity (grey dots) approximately grows linearly in time, and its slope $\gamma$ can be numerically fitted.  Fig.~\ref{fig.slope} (b) depicts the dependence of $\gamma$ versus different $N$ in a log scale. For a small system size $L=10$ and a given tolerance value (grey and orange data), linearly increasing the size of the operator basis seems to exponentially improve the learning quality. A larger system size $L=16$ (blue) notably increases the growth rate in contrast to $L=10$ (grey). This occurs simply because the total Hilbert space enlarges exponentially in system size $L$, and hence it becomes more difficult to reconstruct a quantum state by GD. We also use $\gamma$ to quantify the learning quality for results shown in Fig.~\ref{fig.Violin_main} in the main text.
	
	\subsection{Performance of truncated basis---different tolerance values for ADA-Trotter}
\begin{figure}[h]
		\includegraphics[width=0.6\linewidth]{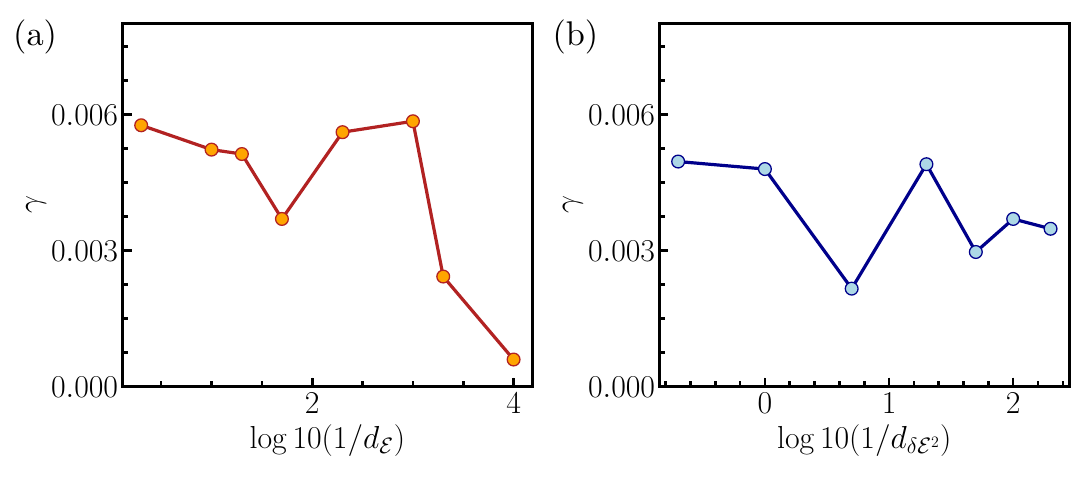}
		\caption{Growth rate $\gamma$ of the optimized state infidelity after GD, for different tolerance values. $\gamma$ always remains small and becomes even smaller for extremely tight tolerances. Although not shown in the plot, we have also verified that the standard deviation of the fitted slope is generally smaller than or comparable to the size of the data points. We use $L=16, J_z=-1,h_x=-1.7,h_z=0.5,$ $N=106$, $M=100$ for numerical simulation and fix $d_{\delta \mathcal{E}^2}=0.01$ in panel (a) and $d_{\mathcal{E}}=0.02$ in (b).
		}
		\label{fig.slope_tolerances}
	\end{figure}

	Now we confirm that for all tolerances considered in this work, learning basis involving $N=106$ operators is sufficient to achieve high-quality QHL. As shown in Fig.~\ref{fig.slope_tolerances} panels (b) and (c), for different energy and energy variance tolerances the fitted slope $\gamma$ is always sufficiently small ($\lesssim 0.006$).  Also, in general the learning quality improves for smaller tolerances. It naturally happens since the ADA-Trotter algorithm tends to use smaller step sizes throughout the entire time evolution, hence the Trotter errors can remain sufficiently local which are earlier to learn.

	\section{Relation to Floquet-Magnus expansion}
	The Floquet-Magnus expansion has been often used to derive effective Hamiltonians for Trotterized systems or periodically driven systems. However, since
	ADA-Trotter adapts the Trotter step, temporal periodicity is violated and the Floquet-Magnus expansion does not apply. However, we note that the effective Hamiltonian obtained by QHL is not completely chaotic. Rather, it exhibits certain interesting connections with the Floquet-Magnus expansion. Apart from the statistical behavior of Trotter errors as shown by the violin plot in Fig.~\ref{fig.Violin_main}, here we directly plot the trajectory of the learned coefficients and compare it with the corresponding Floquet results.

	To do so, we first derive the effective Hamiltonian $H_{\mathrm{eff}}(\tau)$ 
	corresponding to the second order Trotter decomposition $U(\delta t_i) = e^{-i\tau H_x/2}e^{-i\tau H_z}e^{-i\tau H_x/2}
	$ with a fixed step size $\tau$.	
	By employing the expansion for the symmetric BCH expansion $e^C=e^{A/2}e^{B}e^{A/2}$, where $C=A+B-\frac{1}{24}[[B,A],A]-\frac{1}{12}[[B,A],B]$~\cite{casas2009efficient}, we obtain
	\begin{equation}
		\begin{aligned}
			\label{eq.Magnus_expansion_terms}
			H_{\text{eff}} &= H_x+H_z+\frac{\tau^2}{24}[[H_z,H_x],H_x]+\frac{\tau^2}{12}[[H_z,H_x],H_z]+\mathcal{O}(\tau^4)\\
			&=C_X\sum_jX_j+C_Z\sum_jZ_j+C_{ZZ}\sum_jZ_jZ_{j+1}\\
			&+C_{YY}\sum_jY_jY_{j+1}+C_{ZX}\sum_j(X_jZ_{j+1}+Z_jX_{j+1})+C_{ZXZ}\sum_jZ_{j-1}X_jZ_{j+1}+\mathcal{O}(\tau^4),
		\end{aligned}
	\end{equation}
	with the coefficients
	\begin{equation}
		\begin{aligned}
			&C_X= h_x-\frac{2\tau^2}{3}J_z^2h_x-\frac{\tau^2}{3}h_z^2h_x,\ \ 
			C_Z =h_z+\frac{\tau^2}{6}h_x^2h_z,\ \ 
			C_{ZZ}=J_z+\frac{\tau^2}{3}J_zh_x^2,\\
			&C_{YY} = \frac{\tau^2}{3}J_zh_x^2, \ \ C_{ZX}=-\frac{2\tau^2}{3}J_zh_xh_z, \ \ C_{ZXZ}= -\frac{2\tau^2}{3}J_z^2h_x.
			\label{eq.Coefficients}        
		\end{aligned}
	\end{equation}
	Higher order terms are negligible for the small step size considered here.
	
	\begin{figure}[h]
		\includegraphics[width=0.8\linewidth]{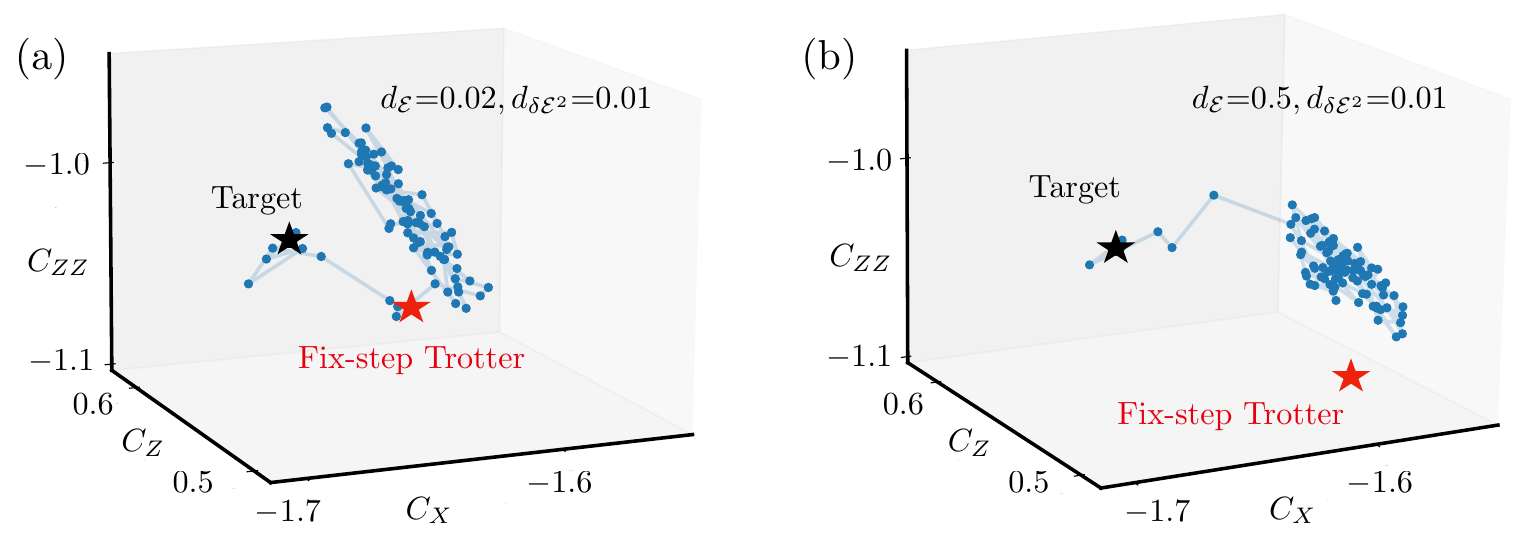}
		\caption{Comparison between trajectories of learned coefficients for ADA-Trotter (blue), fix-step Trotter (red) and the target (black). The cluster of $C_{\alpha}$ seems to be close to the fix-step Trotter prediction. We use $L=16, J_z=-1,h_x=-1.7,h_z=0.5,$ $N=106$,$M=100$ for numerical simulation. Tolerances used for ADA-Trotter are shown in the plot. $\tau=0.198$ for (a) and 0.288 for (b).
		}
		\label{fig.Floquet}
	\end{figure}
	Similar to Fig.~\ref{fig.main} in the main text, here in Fig.~\ref{fig.Floquet} we also show the time trace of the learned coefficients for ADA-Trotter (blue). We calculate the average step size and the corresponding coefficients (red) for fix-step Trotter via Eq.~\ref{eq.Coefficients} for two different sets of tolerances in Fig.~\ref{fig.Floquet} panels (a) and (b). Rather than being completely chaotic, these blue trajectories form a cluster in this three dimensional space and are close to the red star. 
	It may imply that different error terms in the effective Hamiltonian $H_{\text{eff}}^{[t]}$ are correlated with each other, in a similar fashion as in fix-step Trotter algorithm. We leave detailed analysis for the shape of the cluster and their relation with fix-step Trotter error to future works.

\end{document}